\documentclass[conference,a4paper]{APSIPA2021}
\usepackage{graphicx}
\usepackage{amssymb}
\usepackage{threeparttable}
\usepackage{url}
\usepackage{kotex}
\usepackage{xcolor}

\begin{document}

\title{Neural Vocoder Feature Estimation for Dry Singing Voice
Separation}

\author{
\authorblockN{Jaekwon Im, Soonbeom Choi, Sangeon Yong, and Juhan Nam}
\authorblockA{
Graduate School of Culture Technology, KAIST, Daejeon, South Korea \\
E-mail: {\tt \{jakeoneijk,cjb3549,koragon2,juhan.nam\}@kaist.ac.kr}
}
}

\maketitle

\begin{abstract}
Singing voice separation (SVS) is a task that separates singing voice audio from its mixture with instrumental audio. Previous SVS studies have mainly employed the spectrogram masking method which requires a large dimensionality in predicting the binary masks. In addition, they focused on extracting a vocal stem that retains the wet sound with the reverberation effect. This result may hinder the reusability of the isolated singing voice.
This paper addresses the issues by predicting mel-spectrogram of dry singing voices from the mixed audio as neural vocoder features and synthesizing the singing voice waveforms from the neural vocoder.    
We experimented with two separation methods. One is predicting binary masks in the mel-spectrogram domain and the other is directly predicting the mel-spectrogram. Furthermore, we add a singing voice detector to identify the singing voice segments over time more explicitly. 
We measured the model performance in terms of audio, dereverberation, separation, and overall quality. The results show that our proposed model outperforms state-of-the-art singing voice separation models in both objective and subjective evaluation except the audio quality. 

\end{abstract}
\section{Introduction}
Singing voice separation (SVS) is a task of isolating singing voice audio from its musical mixture with various instrumental sounds. SVS is an important topic because the separated singing voice can be used not only in music production, such as music remix, but also in other tasks including singing voice synthesis, singer recognition, lyric recognition, melody extraction and note transcription. Early approaches use methods that take a subspace of singing voice from the decomposed mixed audio, for example, using non-negative matrix factorization (NMF) \cite{nonnegativematrixfactorization} and principal component analysis (PCA) \cite{pca}. Nowadays, deep learning is the dominant approach as it has significantly improved the separation performance \cite{firstunetismir,MonoauralSVS,Open-Unmix,Mmdenselstm,Spleeter}.   

A common processing pipeline in the deep learning approach is to predict the spectrogram masks of the singing voice from the mixed audio spectrogram and multiply the predicted masks with the mixed audio spectrogram element-wise to calculate the magnitude part of the singing voice. The phase part of the singing voice is obtained from that of the mixed audio spectrogram or predicted using a phase reconstruction algorithm such as Griffin-Lim to convert the estimated singing voice spectrogram into a waveform. However, the phase of the mixed audio spectrogram differs from that of the singing voice spectrogram, and the phase reconstruction algorithm also has limitations in predicting the precise phase. In addition, the large dimensionality of spectrogram requires a neural network to predict the masks on the numerous feature bins. Another common practice in the deep learning approach is that the outcome of the separated singing voice is a vocal stem which is processed with audio effects such as reverberation. This mainly attributes to the available multi-track datasets such as MusDB \cite{musdb18} which manage individual sound sources in stem unit for convenience. This wet singing voice with other voice sources may hinder the reuse of the separated singing voice.

Recently, the phase issue has been addressed by predicting the magnitude and phase at the same time in a neural network architecture. For instance, in the Complex as Channel framework (CaC) \cite{CaC}, SVS was evaluated by creating features with magnitude and phase information, taking the real and imaginary parts of the spectrogram as real-valued features, respectively. PhaseNet handled the phase prediction as a classification task by discretizing the phase \cite{phasenet}. ResUNetDecouple predicts complex ideal ratio masks (cIRM) as a way of predicting the information \cite{decoupling}. Nevertheless, it is still challenging to  predict phase and magnitude simultaneously. These studies also have limitations in that dereverberation was not considered.


An alternative to the time-frequency domain masks of singing voice is vocoder features which can be directly converted to waveforms using a pre-built vocoder model. SSSynth estimated the WORLD vocoder parameters (F0, harmonic spectral envelope, and aperiodicity envelope of the singing voice) \cite{world} from mixed audio using a neural network and converted them into the singing voice audio using the vocoder \cite{sssynth}. This method has advantages in that the predicted features are much small dimensional and the separated singing voice has little interference from other instruments. However, F0 has more considerable effect on the audio quality of the synthesized voice in the DSP-based vocoder than other vocoder parameters. SSSynth handles all vocoder parameters uniformly without considering the relative importance of F0. This yields low synthesis quality. Content-based singing voice extraction is another method that predicts the WORLD vocoder parameters of singing voice from mixed audio \cite{sssynth2}. They proposed a model that predicts unprocessed dry vocals not only with accompaniment but also reverberation. However, it has the limitation that the learning process is complex and the predicted quality of the singing voice cannot exceed the synthesis quality of WORLD vocoder. 


In this paper, we propose an SVS model that uses a neural vocoder to address both singing voice separation and dereverberation. The SVS model is composed of two modules. One is a separation module that estimates mel-spectrogram of dry singing voices from the mixed audio as neural vocoder features and the other is the neural vocoder that synthesizes dry singing voice waveforms from the mel-spectrogram. This models has several advantages over previous work. First, mel-spectrogram has smaller dimensions than the regular spectrogram. Therefore, the neural network model for source separation can be more streamlined. Second, since the neural vocoder directly generates waveforms from mel spectrgoram, the phase estimation is not necessary. Lastly, since the separation module and neural vocoder are separately trained, performance can be enhanced by augmenting the train data or modifying the structure in one model while leaving another model unchanged.



This use of neural vocoder in voice separation was recently attempted with application to speech restoration \cite{voicefixer}. However, we explore the approach in singing voice separation along with music audio and also focus on dry voice source. Specifically, we investigate the separation module with two different schemes. One is predicting binary masks in the mel-spectrogram domain and the other is directly predicting the mel-spectrogram. Furthermore, we add a singing voice detector to identify the singing voice segments over time more explicitly. Compared to the wet singing voice with reverberation, the dry singing voice has a significant energy difference between the parts with and without the voice. The singing voice detector facilitates estimating more refined mel-spectrogram by determining the presence of singing voice based on energy. 


We compared our method to SSSynth \cite{sssynth} and ResUNetDecouple \cite{decoupling}. We evaluated each method with objective metrics. However, it is difficult for the audio synthesized by the neural vocoder to be precisely matched with the target audio at the sample level. This makes it difficult to rely on the existing objective evaluation metrics \cite{voicefixer}. Therefore, we evaluated the separation quality, dereverberation quality, and audio quality with more emphasis on subjective evaluation. Sound examples of the compared methods are available online\footnote{\url{https://jakeoneijk.github.io/mel-svs-demopage/}}.



\section{Methodology}

\begin{figure}[t]
\begin{center}
\includegraphics[width=85mm]{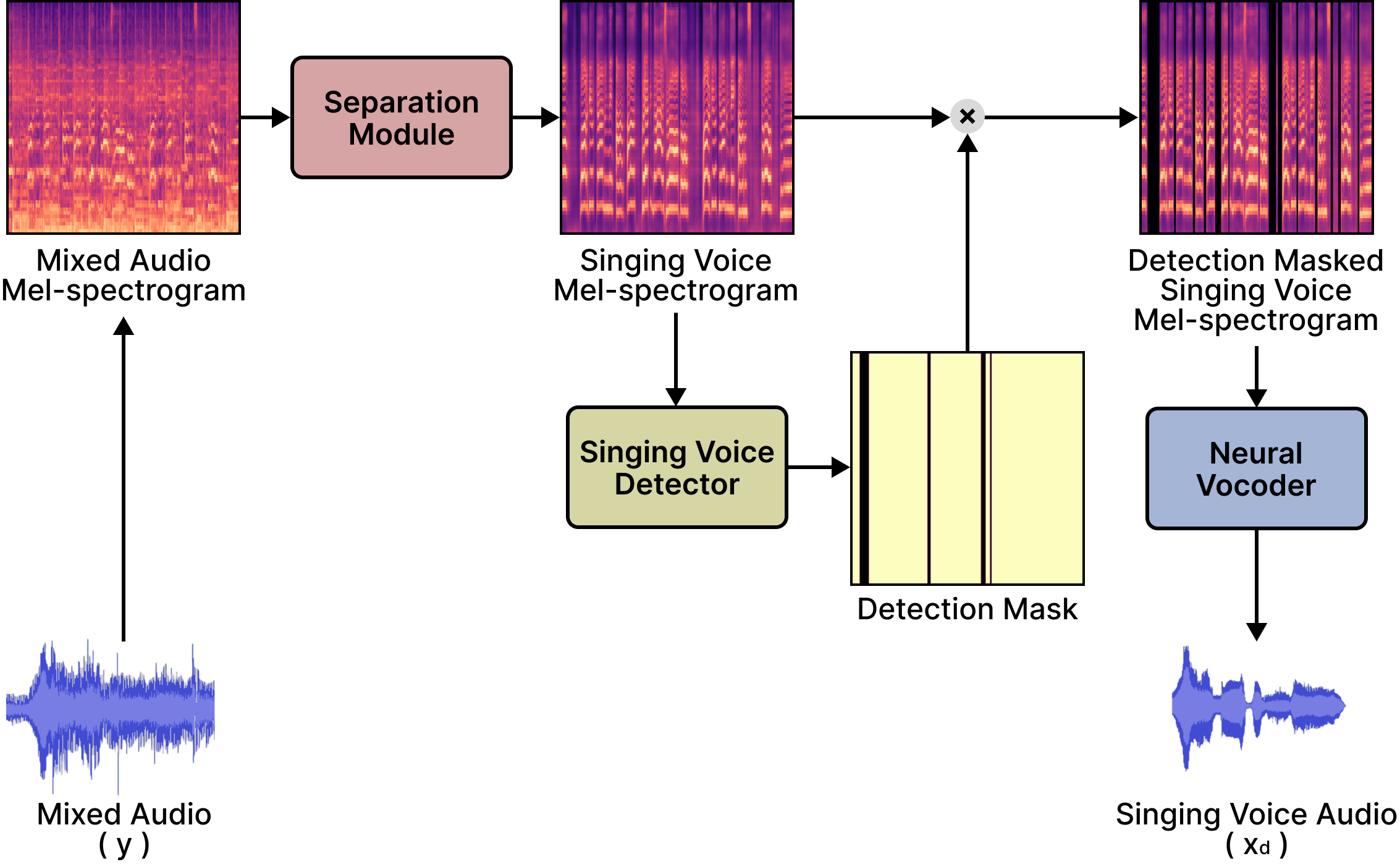}
\end{center}
\caption{The overall architecture of our proposed model.}
\vspace*{-3pt}
\label{fig:overall_architecture}
\end{figure}

We assume 
that mixed audio is the linear sum of dry singing voice, wet singing voice by reverberation, and the accompaniment as follows:
\begin{equation}
 y = x_d + x_r + x_a.
\end{equation}
where $y \in \mathbb{R}^{c \times t}$ is the mixed audio, $x_d \in \mathbb{R}^{c \times t}$ is the dry singing voice, $x_r \in \mathbb{R}^{c \times t}$ is the wet singing voice, and $x_a \in \mathbb{R}^{c \times t}$ is the accompaniment. We define the wet singing voice $x_r$ as the convolution of dry singing voice with a spatial room impulse response (SRIR). 
\begin{equation}
 x_r = \alpha (h \ast x_d).
\end{equation}
where $h \in \mathbb{R}^{c \times t}$ is an SRIR, $*$ and $\alpha$ stands for convolution operation and a reverberation coefficient constant, respectively.

Fig. \ref{fig:overall_architecture} shows the overall architecture. Through short-time Fourier transform (STFT), the mixed audio waveform is turned into the spectrogram. By using the mel filterbank, the spectrogram is transformed into the mel-spectrogram. The separation module estimates the mel-spectrogram of dry singing voice from the mel-spectrogram of mixed audio. From the mel-spectrogram of singing voice ,the singing voice detector predicts the singing voice detection mask, which is a feature that determines whether a singing voice exists at each time frame. Through element-wise multiplication of the detection mask and the singing voice mel-spectrogram, the time-wise masked mel-spectrogram of singing voice is converted to the final waveform audio by the neural vocoder.

\subsection{Separation Module}

\begin{figure}[t]
\begin{center}
\includegraphics[width=86mm]{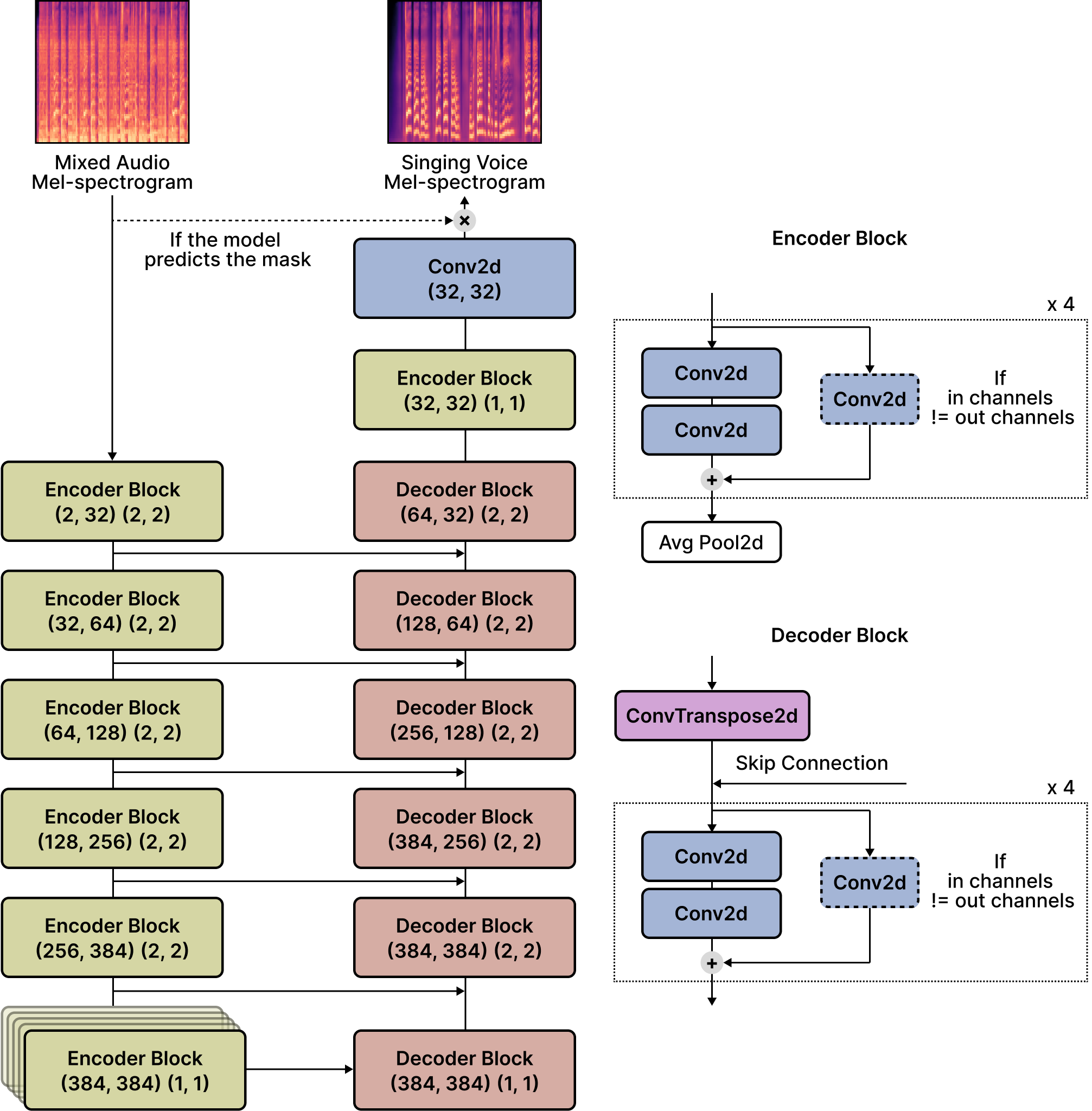}
\end{center}
\caption{Deep ResUNet used in our architecture. The numbers in the block represent (in channels, out channels), (down/upsample ratio).}
\vspace*{-3pt}
\label{fig:separationModule}
\end{figure}

We employ the UNet \cite{firstUnet} architecture for the separation module. UNet has proven to be effective for singing voice separation \cite{firstunetismir}. It is common to set the spectrogram of mixed audio to the input of the encoder and predict the the time-frequency masks of singing voice in the decoder. We designed the configuration of architecture based on ResUNetDecouple \cite{decoupling}. Since we use mel-spectrogram of mixed audio and dry singing voice in the UNet architecture, we modified the modules that face with the input and output. Fig. \ref{fig:separationModule} illustrates the structure of the separation module. Deep Residual UNet is a deep structure in which each encoder block and decoder block consists of 4 residual convolutional blocks, and each residual convolutional layer contains 2 or 3 convolutional layers. All convolutional layers except convolutional layers for skip connection are preceded by batch normalization and application of leaky ReLU. The final output of the decoder becomes a feature with a value between 0 and 1 through the sigmoid activation function. We employed two methods: predicting the mask of the mel-spectrogram and directly predicting the mel-spectrogram. In the case of predicting the mask, the mel-spectrogram of singing voice is calculated by element-wise multiplication of the predicted mask with the mixed audio mel-spectrogram. The separation module is optimized with the mean absolute error (MAE) loss between the predicted dry singing voice and the target dry singing voice.  

\subsection{Singing Voice Detector}
\begin{figure}[t]
\begin{center}
\includegraphics[width=90mm]{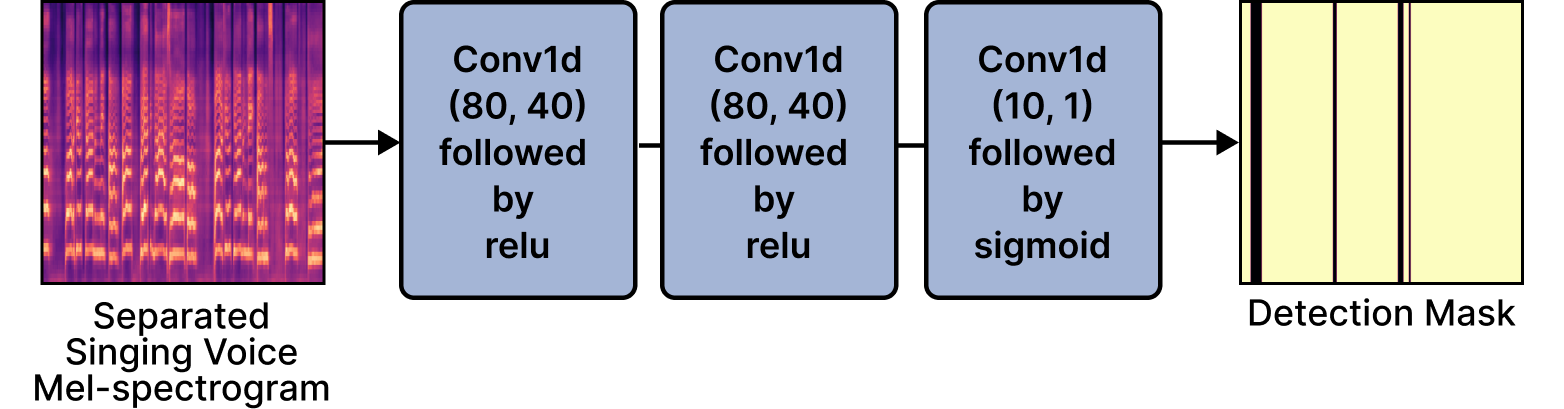}
\end{center}
\caption{Singing voice detector used in our architecture. The numbers in the block represent (in channels, out channels).}
\vspace*{-3pt}
\label{fig:svdetector}
\end{figure}
Dry singing voice has persistent energy only during the voice production while wet singing voice has a decaying energy even after the voice production is paused. As a consequence, determining the presence of singing voice by energy level is simple for dry singing voice. Using this attribute, we design a singing voice detector that predicts a time-wise detection mask with a value of 1 if a singing voice exists and 0 if it does not for each time frame. Fig. \ref{fig:svdetector} shows the structure of the singing voice detector. The singing voice detector is a simple structure with three 1D convolution layers that keep the dimensions on the time axis. The detection mask is created by repeating the predicted 1d feature along the frequency axis according to the frequency dimension size of the mel-spectrogram. The detection mask is multiplied element-wise by the predicted singing voice mel-spectrogram to estimate the masked singing voice mel-spectrogram. The following loss function is used to optimize the separation module with the singing voice detector:
\begin{equation}
    \mathcal{L}_{sep} = ||\hat{X}-X||_{1} + ||\hat{X}_{masked}-X_{masked}||_{1}
\end{equation}
where $X \in \mathbb{R}^{c \times T \times F}$ is the target singing voice, $\hat{X} \in \mathbb{R}^{c \times T \times F}$ is the predicted singing voice, $X_{masked} \in \mathbb{R}^{c \times T \times F}$ is the target masked singing voice, and $\hat{X}_{masked} \in \mathbb{R}^{c \times T \times F}$ is the predicted detection masked singing voice. They are all mel-spectrograms. 


\subsection{Neural Vocoder}
The neural vocoder synthesizes singing voice waveforms from the estimated mel-spectrogram from the separation module. We used HiFi-GAN \cite{kong2020hifi}, a neural vocoder that achieves high quality voice synthesis by using multi-receptive field fusion on a generator, multi-period discriminators and multi-scale discriminators. Since it can synthesize high-fidelity voice in real-time, it is widely used in both speech and singing voice \cite{jeong2021diff, choi2022melody, zhang2022visinger} synthesis in recent years. 



\vspace{5mm}

\section{Experiments}
\subsection{Datasets}
We used MUSDB18 which is one of the most widely used datasets for SVS \cite{musdb18}. Since our experiment requires dry monophonic singing voices, however, the MUSDB18 dataset offering wet polyphonic singing voices is not appropriate. Therefore, we collected dry singing voices of songs in the MUSDB18 dataset from MedleyDB\cite{Medleydb} and the website: Mixing Secrets for The Small Studio\footnote{\url{https://www.cambridge-mt.com/ms/mtk/}}, which are original sources of the MUSDB18 dataset. The MUSDB18 dataset has 86 songs for training, 14 songs for validation, and 50 songs for testing. We trained the separation module with the accompaniment of the MUSDB18 dataset and the corresponding dry singing voice. The neural vocoder was also trained with dry singing voices of the same dataset. We used the same 86 songs for training the separation for training HiFiGAN. We used the v1 setting for HiFi-GAN while changing the sampling rate to 24,000 Hz to match with the separation module output.

We used the DetmoldSRIR dataset \cite{detmoldsrir} to investigate various reverberation conditions. The dataset provides SRIRs measured in three performance spaces at the Detmold University of Music. The first room is the Detmold Konzerthaus (medium-sized concert hall), the second is the Brahmssaal (small music chamber room), and the last is the Detmold Sommertheater (theater). We randomly divided 937 SRIRs into a train, valid, and test set to contain 749, 94, and 94 SRIRs, respectively. In the training stage, the SRIRs were randomly selected at every step. For the test, we made 94 pairs of music and SRIR by matching 1 or 2 songs per SRIR.

\subsection{Data Processing}
For the convenience of experiments, all audio sample rates were resampled to 24,000 Hz. We adopted the mix-audio data augmentation \cite{mixaudioaugmentation} to increase the volume of the training set. The mix-audio data augmentation a method that generates training data by taking each source from different music tracks. We performed the mix-audio data augmentation using two accompaniments and one singing voice. Since the vocoder synthesize a monophonic voice, only the situation with one singing voice was considered. All audio files used in the training phase was randomly segmented by 3 seconds. The reverberation of the singing voice was rendered by the convolution operation between randomly selected SRIR and dry singing voice sources. The reverberation coefficient constant was randomly set to a value between 0 and 1 every time.

The audio was converted to the spectrogram by short-time Fourier transform (STFT) using 1024 samples in FFT size, 256 samples in hop size, and the Hann window function. The spectrogram was converted to mel-spectrogram through 80 channel mel filterbank. The log magnitude compression was applied to the mel-spectrogram. When the separation module was trained to directly estimate the mel-spectrogram, the min-max normalization was applied to set the value of the mel-spectrogram between 0 and 1.

The target detection mask was obtained in the following steps. First, the spectrogram was squared element-wise. Second, it is summed over the frequency axis to obtain an one-dimensional feature. Third, values less than a threshold are mapped to 0 and the others are mapped to 1. 
In our experiment, the threshold was set to 4. All training was executed with a batch size of 4 and the Adam optimizer \cite{adam}. The learning rates were set at 0.001 and decreased by a rate of 0.9 every 15000 steps. Each separation module was trained by 1 million steps.

\subsection{Compared Models}
We compared our method with SSSynth and ResUNetDecouple. SSSynth estimates the World vocoder parameters of singing voice from mixed audio and ResUNetDecouple estimates the spectrogram domain masks. We trained the two models using the same dataset, data split and preprocessing explained above.   

\subsection{Objective Evaluation}
Since the audio synthesized by the neural vocoder does not perfectly match the target at the sample level, the conventional source separation evaluation metrics such as source to distortion (SDR), source to interference (SIR) ratios \cite{sdr} is not appropriate. We verified that even the singing voice synthesized from the target mel-spectrogram has significantly low values of SDR  although it perceptually has high-quality. To address this issue, we used two spectral metrics. First, we used scale-invariant spectrogram to noise ratio (SiSPNR) \cite{voicefixer}. SiSPNR is a spectrum-domain metric similar to the scale-invariant signal to noise ratio (SiSNR) \cite{sisnr}, comparing the energy of a signal and its background noise.
The formula of SiSPNR is as follows.

\begin{equation}
SiSPNR = 10 \log_{10}{\frac{||S_{target}||^2}{||e_{noise}||^2}}
\end{equation}
\begin{equation}
S_{target} = {\frac{<\hat{S},S>S}{||\hat{S}||^2}}
\end{equation}
where $\hat{S}$ is the magnitude spectrogram of the predicted signal, and $S$ is the magnitude spectrogram of the target signal. As a second metric, we propose a spectrogram to Distortion Ratio (SPDR). SPDR is the same as signal-to-distortion ratio(SDR) \cite{sdr}, except that it is calculated with the magnitude spectrogram. The formula of SPDR is as follows.

\begin{equation}
SPDR = 10 \log_{10}{\frac{||S||^2}{||e_{interf}+e_{noise}+e_{artif}||^2}}
\end{equation}
where $S$ is the magnitude spectrogram of the target signal.

\subsection{Subjective Evaluation}
We also conducted subjective evaluation through a listening test with 23 participants. Considering the diversity of music genre and reverberation type, 15 songs from the test set were selected for the listening test. For each song, the singing voices predicted by SSSynth, ResUNetDecouple, the proposed method without and with the singing voice detector were evaluated. We only evaluated the audio estimated by predicting the mask for the efficiency of evaluation because it has no significant performance difference from directly predicting the mel-spectrogram. We also included the target audio in the evaluation audio set as the upper bound of the result. The evaluation items included audio quality, dereverberation performance, separation performance, and overall quality. Each evaluation item was rated from 1 to 5 points. 


\vspace{5mm}

\section{Results}


\subsection{Evaluation Results}

\begin{table}[t]
\centering
\begin{threeparttable}
\caption{
SPDR and SiSPNR comparison of previous and our proposed methods.}
\label{objectiveeval}
\begin{tabular}{|l|l|l|}
\hline
                                                                    & SPDR↑  & SiSPNR↑    \\ \hline
Target                                                              & -     & 111.81    \\ \hline
HiFi-GAN using target mel\tnote{a}                                  & 13.41 & 10.17     \\ \hline
SSSynth                                                             & 7.53  & 3.52      \\ \hline
ResUNetDecouple                                                     & 8.66  & 4.87      \\ \hline
Proposed (directly predict mel\tnote{a}\hspace{0.4em}) w/o SVD\tnote{b}\quad  & 9.94  & 5.95      \\ \hline
Proposed (directly predict mel\tnote{a}\hspace{0.4em}) w/ SVD\tnote{b}\quad   & 10.19 & 6.30      \\ \hline
Proposed (predict mel\tnote{a}\hspace{0.4em} mask) w/o SVD\tnote{b}\quad      & 10.13 & 6.29      \\ \hline
Proposed (predict mel\tnote{a}\hspace{0.4em} mask) w/ SVD\tnote{b}\quad       & \textbf{10.35} & \textbf{6.43}\\ \hline
\end{tabular}
\begin{tablenotes}
    \item[a] mel-spectrogram
    \item[b] Singing voice detector
\end{tablenotes}
\end{threeparttable}
\end{table}

Table \ref{objectiveeval} shows the SiSPNR and SPDR comparison of the proposed and two compared methods. SiSPNR and SPDR results reveal similar tendencies. In two evaluation metrics, our proposed method outperforms the previous methods. The method of directly predicting the mel-spectrogram performed lower than the method of predicting the mask of the mel-spectrogram. It implies that the advantage of predicting the mask also applies to SVS in the mel-spectrogram domain. The results show that the singing voice detector can effectively improve the performance of the separation module. The performance of the models with the singing voice detector is superior to that of the models without the singing voice detector.

\begin{table}[t]
\centering
\begin{threeparttable}
\caption{MOS score comparison of previous and our proposed methods.
}
\label{subjectiveeval}
\begin{tabular}{|l|c|c|c|c|}
\hline
                 & \begin{tabular}[c]{@{}c@{}}Audio \\ Quality\end{tabular} & \begin{tabular}[c]{@{}c@{}}Dereverb\tnote{e} \\ Performance\end{tabular} & \begin{tabular}[c]{@{}c@{}}Separation \\ Performance\end{tabular}  & \begin{tabular}[c]{@{}c@{}}Overall \\ Quality\end{tabular} \\ 
\hline
Target              & 4.81 ± 0.49 & 4.61 ± 0.82 & 4.78 ± 0.62 & 4.79 ± 0.47 \\ \hline
Model1 \tnote{a}    & 1.32 ± 0.65 & 2.18 ± 1.30  & 1.94 ± 1.14 & 1.43 ± 0.73 \\ \hline
Model2 \tnote{b}    & \textbf{3.41} ± 0.89  & 3.29 ± 1.05 & 2.53 ± 1.04 & 3.19 ± 0.78\\ \hline
Model3 \tnote{c}    & 3.07 ± 0.86 & 3.48 ± 2.44 & 3.64 ± 1.02 & 3.24 ± 0.82\\ \hline
Model4 \tnote{d}    & 3.17 ± 0.84 & \textbf{3.50} ± 1.02 & \textbf{3.71} ± 1.0 & \textbf{3.35} ± 0.84\\ \hline
\end{tabular}
\begin{tablenotes}
    \item[a] SSSynth.
    \item[b] ResUNetDecouple.
    \item[c] The proposed method without Singing Voice Detector.
    \item[d] The proposed method with Singing Voice Detector
    \item[e] Dereverberation
\end{tablenotes}
\end{threeparttable}
\end{table}

Table \ref{subjectiveeval} shows the Mean Opinion Score (MOS) results of previous and proposed methods. Our proposed methods scored better than previous methods in all areas except for audio quality. In particular, our methods significantly outperformed the previous methods in the separation performance. The reason our methods scored lower than ResUNetDecouple in audio quality is presumably due to artifacts caused by HiFi-GAN. These artifacts can be alleviated by training HiFi-GAN with data with more diverse distributions of pitch and timbre. On every metric, the model with the singing voice detector outperformed the model without it. This indicates that the singing voice detector is an effective module for dry singing voice separation. 

\subsection{Spectrogram analysis}

\begin{figure}[t]
\begin{center}
\includegraphics[width=80mm]{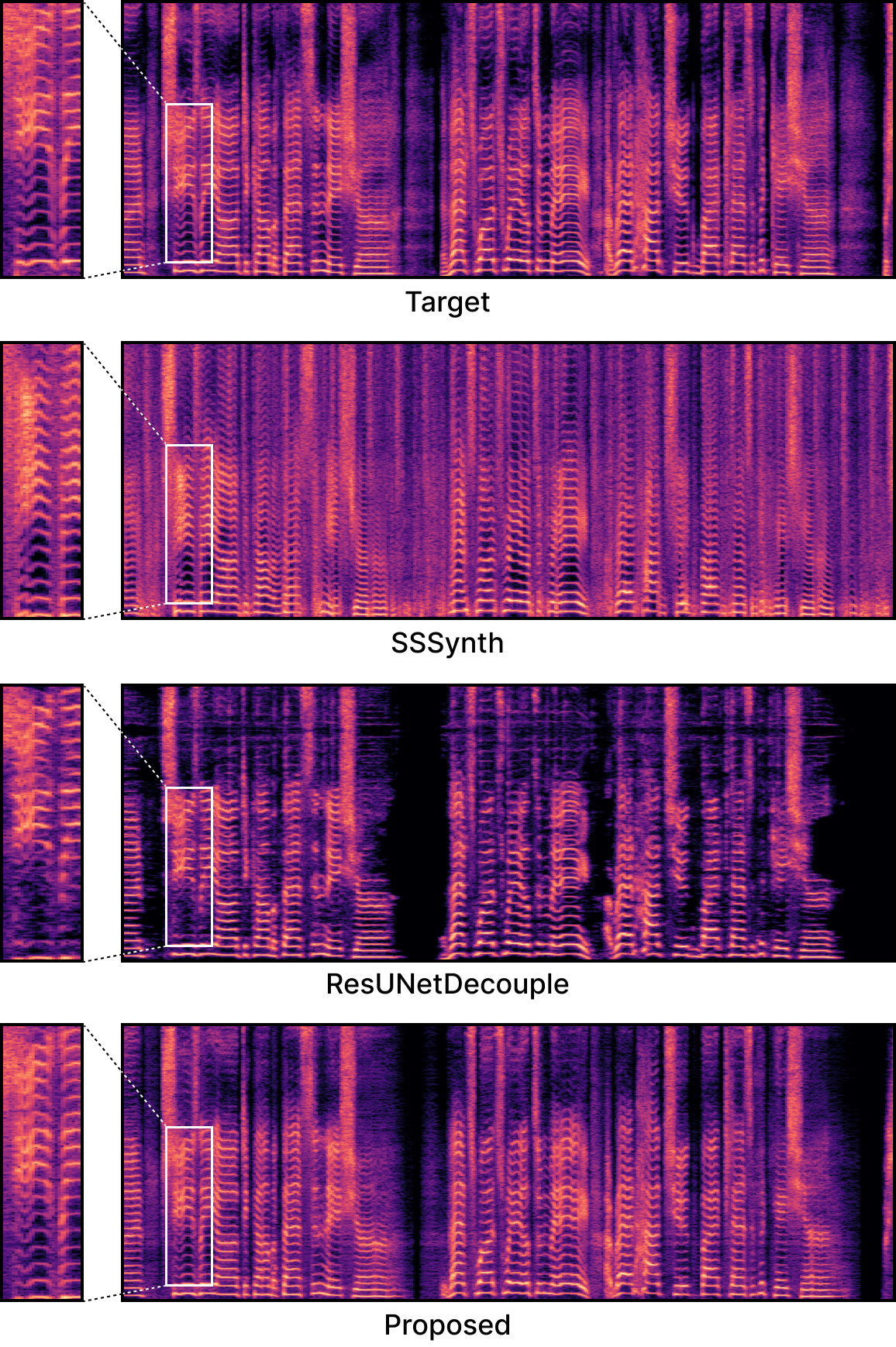}
\end{center}
\caption{The target (ground truth) and estimated singing voice from the three different methods. The proposed model used the setting that predicts mel-spectrogram masks with the singing voice detector.}
\vspace*{-3pt}
\label{fig:spectrogramanalysis}
\end{figure}

We compared the singing voice separation methods through spectrogram analysis to provide more insightful ideas about the results. Fig. \ref{fig:spectrogramanalysis} depicts the spectrograms of the separated singing voice estimated by SSSynth, ResUNetDecouple, and the proposed model. 
SSSynth tends not to generate the overall harmonic structure of sound correctly. Since it is difficult to predict the f0 of a singing voice with reverberation from the mixed audio and the world vocoder is greatly affected by incorrect F0, the singing voice separated by SSSynth has many artifacts. Our proposed model outperforms ResUnetDecouple in predicting the harmonic structure of the singing voice. In addition, we found that our model separates percussion sounds better than ResUnetDecouple. This is probably because our proposed structure is highly optimized for predicting dry singing voices. Since the neural vocoder is optimized to synthesize solely dry singing voices, other instrumental sounds are not likely to be generated. Furthermore, since the singing voice detector helps the separation module be trained regarding the existence of the singing voice, the separation module can effectively separate the instrumental sounds when the singing voice is absent.

\section{Conclusions}
In this study, we proposed an SVS model that estimates neural vocoder features for dry singing voice separation. In objective and subjective evaluation, our methods showed better performance than the compared methods. In addition, we proposed the singing voice detector to improve the dry singing voice separation and verified its effectiveness through the objective and subjective evaluation. In the future, we plan to improve the audio quality of singing voice with more enhanced separation modules and neural vocoders. 

\bibliographystyle{unsrt}
\bibliography{APSIAP2022.bib}

\end{document}